\documentclass[useAMS,usenatbib,a4paper]{mn2e}
\usepackage{graphicx}% This allows you to call \includegraphics

\def\etal{{\it et al.}\thinspace}
\def\ie{{\it i.e.,}\thinspace}
\def\eg{{\it e.g.,}\thinspace}

\def\eq{\begin{equation}}
\def\en{\end{equation}}

\def\etal{{\it et al.}\thinspace}
\def\ie{{\it i.e.}\thinspace}

\def\eg{{\it e.g.}\thinspace}
\def\apj{{\it Ap.J.}\thinspace}
\def\apjl{{\it Ap.J. Lett.}\thinspace}
\def\apjs{{\it Ap.J. Suppl.}\thinspace}
\def\aap{{\it A\&A}\thinspace}
\def\mnras{{\it MNRAS}\thinspace}
\def\P3hat{{\mathaccent 94 P}_3}

\def\nat{{\it Nature}\thinspace}

\title[The `periodic nulls' of radio pulsar J1819+1305]{The `periodic nulls' of radio pulsar J1819+1305}
\author[Joanna M. Rankin \& Geoffrey A.E. Wright]{Joanna M. Rankin $^{1}$\thanks{Joanna.Rankin@uvm.edu;G.Wright@sussex.ac.uk} and Geoffrey A.E. Wright $^{2}$\footnotemark[1]\\
$^{1}$Physics Department, University of Vermont, Burlington, VT 05405\\
$^{2}$Astronomy Centre, University of Sussex, Falmer, BN1 9QJ, UK}

\begin{document}

\date{}

\pagerange{\pageref{firstpage}--\pageref{lastpage}} \pubyear{2006}

\maketitle

\label{firstpage}

\begin{abstract}
We present a single-pulse study of the four-component pulsar J1819+1305, 
whose ``null'' pulses bunch at periodic intervals of around 57 times the rotation 
period.  The emission bursts between the null bunches exhibit characteristic 
modulations at two shorter periodicities of approximately 6.2 and 3 times the 
rotation period, the former found largely in the two outer components, and 
the latter only in the first component.  Many bursts commence with bright 
emission in second component, exhibit positive six-period drift across the full 
profile width, and end with 3-period modulation in the leading component.  
The 57-period cycle can be modelled geometrically as a sparsely filled 
subbeam carousel with nulls appearing whenever our line of sight intersects 
a circulating empty region.  This interpretation is compatible with other 
recent evidence for periodic, carousel-related nulling and appears to support 
the physics of a polar-gap emission model for ``drifting'' subpulses, but the 
subtle structure of the emission bursts defies an easy explanation.
\end{abstract}
\maketitle

\section{Introduction}

Pulsar J1819+1305 was discovered twice, once in a Swinbourne survey 
(Edwards \etal\ 2001) and also in the Arecibo 430-MHz survey of Navarro 
\etal (2003, hereafter NAF).  The latter paper notes that this pulsar is the 
brightest of the 9 stars discovered, and using pulse-sequence (hereafter 
PS) observations they were able to identify several of its outstanding 
characteristics.  First among these is the star's tendency to ``null'' roughly 
half of the time, a feature not uncommon in pulsars with a large spin-down 
age (47 Myr for J1819+1305). However, a second very strange circumstance 
is that the observed ``null'' intervals exhibited a strong periodicity of 53$\pm$3 
times the rotation period (hereafter $P_1$) of 1.06 s.  A 1000-period section 
of their observation is shown in their fig. 6, and some 18 nearly equally 
spaced bursts of emission are clearly evident.  The figure also shows the 
star's asymmetric, apparently triple, profile and the text notes that a part of 
the intensity variation is due to ``strong intensity modulation of the first 
component''.   

\begin{table}
\label{tab1}
 \begin{center}
 % \begin{minipage}{140mm}
   \caption{Arecibo Single-Pulse Polarimetry Observations}
   \begin{tabular}{ccccc}
   \hline
 Band  & MJD  &  chans & Resolution & Pulses  \\
            &  Date &   &      (\degr)      & RFI   \\
   \hline
   P      &       52707$^*$    &  128 & 0.35 & 1697 \\
            & 2003 Mar 10&       &          &  clean \\
   P      &       52941    & 256  & 0.17 &   955 \\
            & 2003 Oct 29 &       &          &  bad \\
   P      &       53378    & 512  & 0.44 &   848 \\
            & 2005 Jan 8 &       &           &   good \\
   P      &       53778    & 256  & 0.51 &  3394 \\
            & 2006 Feb 12 &       &          &  clean \\
   L      &       53859    &  64  & 0.35 &   2075 \\
            & 2006 May 4 &       &           &   poor \\
 \hline
 \end{tabular}
 \end{center}
 {\small $^*$Unfortunately, an error in the WAPP software 
 resulted in only one linear polarization being recorded, 
 four-fold redundant, during the initial 2003 March session.} 
 %\end{minipage}
 \end{table}

The J1819+1305 section of the NAF paper remarks that little further 
analysis could be made without polarimetry and thus an understanding of 
the pulsar's emission geometry, but that, were this available, ``a proper 
interpretation of the nulling might lead to new insights''.  We have observed 
the pulsar a number of times with the intention of studying its periodic nulling 
behaviour. Our observations are described in \S 2, and \S 3 assesses the 
star's fundamental emission geometry.  \S 4 presents our analysis of the 
``nulls'' and their periodicity and \S 5 analyses the modulation of the emission 
bursts. In \S 6 we interpret the null periodicity as a ``carousel'' circulation time 
and give examples of the often incomplete patterns which can reproduce the 
pulsar's observed behaviour.  \S 7 summarises and discusses our results. 

\begin{figure}
\begin{center}
\includegraphics[width=8cm]{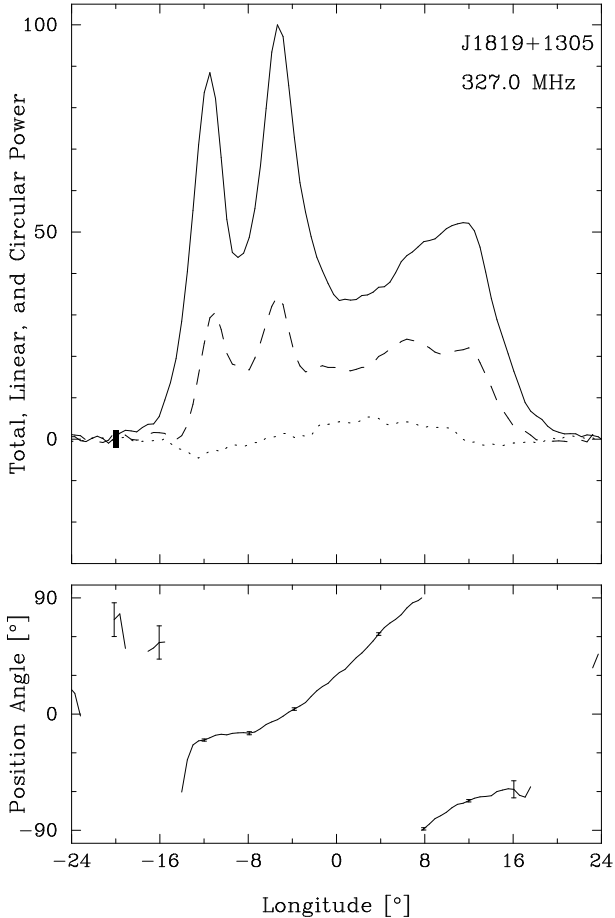}
\caption{Integrated profile of 3394 pulses from J1819+1305 at 327 MHz.  
The top panel gives the the total intensity (Stokes $I$; solid curve), the 
total linear (Stokes $L$; dashed), and the circular polarization (Stokes 
$V$; dotted).  The lower panel shows the average polarisation-angle 
(hereafter PA) traverse.}
\label{fig1}
\end{center}
\end{figure}

\section{Observations}
The observations used in our analyses were made using the 305-m Arecibo 
Telescope in Puerto Rico.  The 327 (P band) and 1400-MHz (L band) polarized 
PSs were acquired using the upgraded instrument together with the Wideband 
Arecibo Pulsar Processor (WAPP\footnote{http://www.naic.edu/$\sim$wapp}) 
over the last three years.  The auto- and cross-correlations of the channel voltages produced 
by receivers connected to orthogonal linearly polarized feeds (circular after 2004 
October 11) were 3-level sampled.  Upon Fourier transforming, sufficient channels 
were synthesized across a 25-MHz (100-MHz at L band) bandpass to provide 
resolutions of some 1 milliperiod of longitude.  The Stokes parameters have been 
corrected for dispersion, interstellar Faraday rotation,\footnote{The star's rotation 
measure was measured as +119$\pm$4 rad-m$^2$} and various instrumental 
polarization effects.  At L band, four 100-MHz channels were observed with centres 
at 1170, 1420, 1520 and 1620 MHz.  Some of the observations encountered 
virtually no interference (hereafter RFI);  others, however, were rendered useless 
for some analytical purposes by radar and other RFI sources.  We were able to 
largely clean the several brief RFI events from the MJD 53778 observation, but 
for two others (MJDs 52941 and 53859) this has not been fully possible.

%sJ1819+1305.52707p (P1639+mar)  1697 pulses  0.352 deg/smpl
%	clean	56.9 P1
%	show fft	58.6+/-1.0		not regular pattern
%sJ1819+1305.52941p  P1639  955 pulses 0.174 deg/smpl
%	triangular pedestal??   Lots of withering interference!!
%	no meaningful measurement possible
%sJ1819+1305.53378p  P1978  848 pulses  0.435 deg/smpl
%	one big spike at abt #50  56.9 P1  nearly periodic nulls
%	hrf  57.3+/-1.0 cleaned PS	lrf  57.3+/-1.0 cleaned
%sJ1819+1305.53778p P2110  3394 pulses 0.510 deg/smpl
%   	clean	51.2 P1 partial LRF  56.95+/-1 HRF
%sJ1819+1305.53859l  P2110  2075 pulses 0.348 deg/smpl
%	Bands 1&2 clean positively, but bad negative-going RFI in all bands

\section{Emission Geometry}
The needed polarisation profile of J1819+1305 is given in Figure~\ref{fig1},  
and here we again see the star's asymmetric three-componented profile.  
Both recent P-band observations exhibit nearly identical such profiles, and 
those at L-band are also quite similar, apart from being more compact and 
having a somewhat more intense trailing ``component''.  The profiles from 
both bands show the prominent, positive-going polarisation-angle (hereafter 
PA) traverse with a central slope $R$ of about +7\degr/\degr.  

Overall, the profile appears entirely conal, and the linear polarisation shows 
evidence of a fourth component which is echoed by inflections in the total 
power.  Longitude-longitude correlation maps at zero delay (not shown) 
exhibit a roughly symmetrical form with distinct regions of self-correlation 
near $\pm$10\degr and a third broad area between some $\pm$7\degr.  
The conal character of J1819+1305's emission is supported by its small 
value of $B_{12}/P_1^2$ (where $B_{12}$ is the surface magnetic field in units 
of 10$^{12}$ Gauss) of 0.6, reflecting its nearly 1-s $P_1$ and small spindown.  
This interpretation makes the pulsar a rare member of the conal quadruple 
(cQ) class (\eg, see Rankin 1993a,b; hereafter Paper VI), where our sightline 
cuts two concentric emission cones but fails to encounter appreciable core 
radiation.  Note also the pronounced edge depolarisation, which is a reliable 
demarkation of the outer conal edge (Ramachandran \& Rankin 2003) due 
to the presence of both orthogonal polarisation modes (OPMs) at roughly 
equal intensity.  The pulsar is too weak for these to be seen in individual 
pulses, but each of our average profiles gives evidence of the ``90\degr\ 
jumps'' on one or both edges that they often produce.  It is interesting to 
compare J1819+1305's profile with that of the long known B1918+19 (\ie, 
Hankins \& Wolszczan 1987: Brown \etal\ 2007) which has a similar 
sightline geometry and thus profile, but a conal triple (cT) form.  In both 
cases the trailing half of the profile is attenuated, perhaps a result of 
``absorption''. 

On this basis it is straightforward to work out the star's basic geometry.  
The P and L-band profiles have outside half-power widths of some 28\degr\  
and 23\degr, respectively, providing a decent estimate of its 1-GHz width 
$\Delta\Psi$ of 24\degr.  Then following the procedures of Paper VI, 
J1819+1305's magnetic latitude $\alpha$ and sightline-impact angle 
$\beta$ can be computed from $R$ and $\Delta\Psi$ to be some 22\degr\ and 
+3.0\degr, respectively---making $\beta/\rho$ 0.53 for the outer cone, 
where $\rho$ is the emission-cone radius to the outside half-power point.  
Moreover, the roughly 15\degr\ outside, half-power width of the inner 
conal component pair squares well with this value of $\alpha$.  Given 
he clear flattening of the PA traverse at the profile edges, we take the 
sign of $\beta$ to be positive.

\begin{figure}
\begin{center}
\includegraphics[width=63mm,angle=-90]{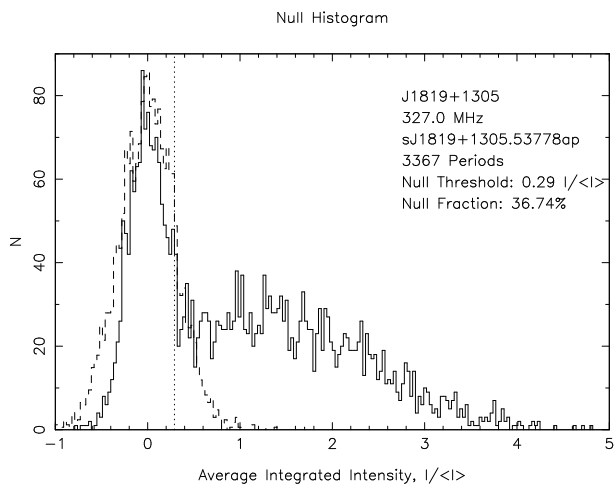}
\caption{Null histogram corresponding to the long 327-MHz observations 
of MJD 53778 (as in Fig.~\ref{fig1}).  The solid and dashed curves 
represent the pulse and off-pulse regions, respectively.  The pulse- 
and null-intensity distributions clearly overlap, and the vertical dotted 
line at 0.29 $<$$I$$>$ shows an arbitrary, but conservative boundary 
between the two populations.  On this basis, some 39\% of B1819+1305's 
pulses are ``nulls''.}
\label{fig2}
%This figure needs some work.  Why is the pulse distribution 
%	around zero intensity so much broader than the off-pulse
%	distribution??  Is this an artefact of the way Stephen's 
%	baseline cleaning code works?
\end{center}
\end{figure}

\begin{figure}
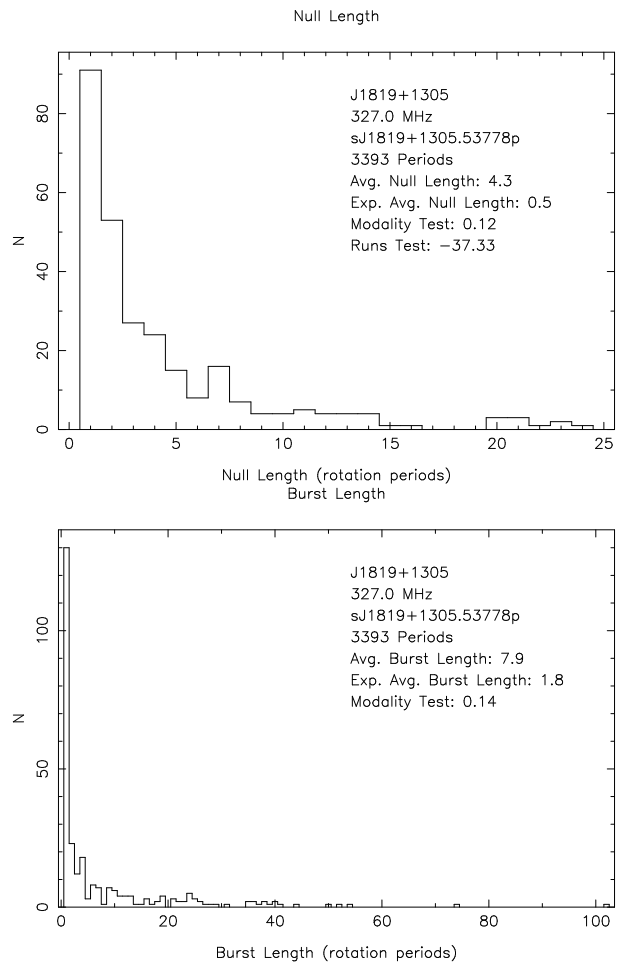

\begin{center}
\includegraphics[width=63mm,angle=-90]{MN-07-1367-MJr1_fig3a.ps}
\includegraphics[width=63mm,angle=-90]{MN-07-1367-MJr1_fig3b.ps}
\caption{``Null'' and burst-length distributions computed from the same 
PS as in Figs.~\ref{fig1} and \ref{fig2}.  Both burst and ``null'' lengths of 
1 $P_1$ appear to be favored, and in both cases the mean lengths 
are 5-10 $P_1$.  Significantly, however, the maximum duration of 
both the bursts and ``nulls'' seems to be around 50 $P_1$.}
\label{fig3}
\end{center}
\end{figure}

%\begin{figure}
%\includegraphics[width=63mm,angle=-90]{prof_shapes_nulls.ps}
%\caption{Total profile together with average profiles of pulses 
%just prior to, immediately following or ``nulls''---again from the 
%same PS as in Fig.~\ref{fig1}.  Pulses just before nulls tend to 
%be slightly early, especially in the bright second component, 
%and the ``null'' profile does show weak power at the positions 
%of the bright leading components.}
%\label{fig4}
%\end{figure}

\begin{figure}
\begin{center}
\includegraphics[width=63mm,angle=-90]{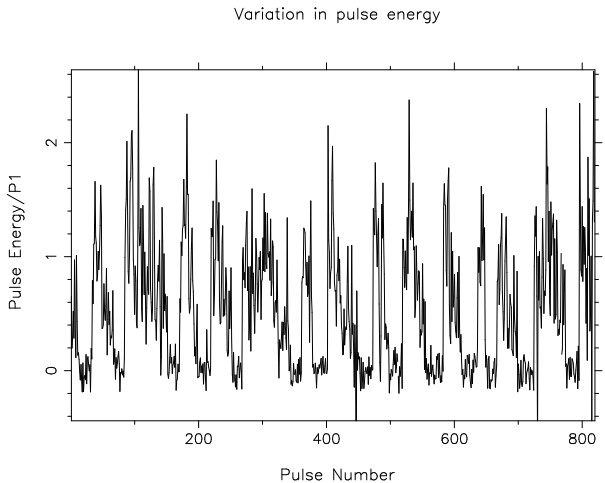}
\caption{Total intensity {\it vs.} pulse number for the 327-MHz PS 
of MJD 53378.  Note the ``periodic nulls'' which though not fully  
regular are more nearly so in this PS than others.  The last part 
of the PS was corrupted by RFI (as are the negative deflections 
near pulses 440 and 725) and so is omitted.}
\label{fig4}
\end{center}
\end{figure}

\section{Analysis}
\subsection{The null and emission distribution}
%both burst and null mean lengths remarkably short wrt 50 P
%	for nulls, 1-P most likely---quite possibly partially due to conflation of pulses & nulls
%	however, no nulls longer than abt 25 P; whereas bursts seen up to abt twice this long
%	last before and first after profiles closely resemble total profile; suggestion that 
%		both are broader, and last before may be slightly earlier;  null average not
%		completely noise-like--sig. excess power at longitudes of first two comps.

Although B1819+1305 is reasonably strong at 327 MHz, its null histogram 
in Figure~\ref{fig2} shows that its nulls and weak pulses cannot be fully 
distinguished, if indeed they are distinguishable at all. The pulse-intensity 
distribution is sufficiently broad that some apparently null pulses may simply 
correspond to pulses in the weak tail of the distribution. Only the large 
secondary distribution around zero indicates that there is a distinct population 
of ostensibly ``true'' nulls whose intrinsic properties make them indistinguishable 
from weak normal pulses.

%Finally, Figure~\ref{fig4} compares the total profile form with partial averages 
%comprised of those pulses just before, just after and during nulls.  All three 
%show power at the longitudes of the two bright leading components, and the 
%first two show softer edges indicating increased power at longitudes outside the profile %edges. 
%There is also the suggestion that comp. 2 arrives a 
%little earlier on average in pulses just before nulls.

%\begin{figure}
%\includegraphics[width=73mm,angle=-90]{PQsJ1819+1305.53378p_lrf.ps}
%\caption{Longitude-resolved fluctuation spectra of the MJD 53378 
%PS in Fig.~\ref{fig5}.  The prominent 50-$P_1$ feature is seen in 
%each of our PSs, and is the only response seen in average spectra 
%of 500 or more pulses.  Note also that the 50-$P_1$ modulation is 
%seen over the full width of the star's profile.  A 256-length FFT was 
%used to produce these spectra.}
%\label{fig6}
%\end{figure}

In order to proceed, we took a conservative definition of the intensity boundary 
between the pulses and ``nulls'', 0.29 $<$$I$$>$, shown by a vertical dotted 
line in the histogram.  On this basis, some 1/3 of the pulses are ``nulls''.  We 
then identified the lengths of contiguous ``nulls'' and ``bursts'' in our 
3392-length PS at 327 MHz on MJD 53778, and the resulting distributions 
are shown in Figure~\ref{fig3}.  In the upper panel we see the null-length 
distribution apparently dominated by the 90 1-period nulls, a total which 
statistically declines to just 4 for 10-period nulls in a manner roughly 
conforming with a random distribution among the non-null pulses. However, 
this distribution comprises only 780 or 2/3 of all the ``nulls", still leaving some 
420 non-random ``nulls" which are contained in sequences of 10 periods or 
more: note the cluster of 10 null sequences with lengths between 20 and 24 
$P_1$. A consistent interpretation of this result would be to view the short 
``nulls" as predominantly weak normal pulses falling randomly within the 
emission bursts, with the ostensible ``true" nulls occurring in clusters of 10 
or more.  Allowing for an overlap between these distributions, one may 
therefore estimate that no more than about 10\% (1/3 of 1/3) of all pulses in 
J1819+1305 could be ``true" nulls.\footnote{The RUNS test result in the 
figure also indicates that the nulls are greatly over-clustered; see Redman 
\& Rankin (2007).}

%NB a test of this might be to separately integrate all the nulls in sequences of > or < 10 pulses.

%Geoff, now that we have the statistics on the plots, is it possible that the 
%		text both below and above can be shortened somewhat per the referee?

In the burst-length distribution (Fig.~\ref{fig3}, lower panel), the total of 130 single-pulses 
preceded and followed by a null may seem high, but in fact this is well below 
what would be expected from a random distribution of nulls with 1/3 chance among 
non-null pulses with a 2/3 probability.  This expectation would be approximately 
$(0.33)^2*0.66*3392 = 250$ pulses. To obtain the observed distribution, the 
chance of a null occurring before or after a non-null pulse must be closer to 1/4, 
and for a 2-pulse burst just 10\%, reflecting the observed non-random bunching 
of the non-null pulses.  The implication is that the probability of a null after a null 
or series of nulls must be considerably higher than 1/3, thus encouraging the 
formation of long ($> 10$ $P_1$) nulls which comprise the ``periodic'' null bunches. 
However, as with the null histogram, note that some very long emission sequences 
are observed---two over about 75 and four between 40 and 60 periods---showing 
that the apparent periodicity of the null bunches is not a precise rhythm.

\subsection{The slow ($\approx$ 50 $P_1$) modulation}
As reported by NAF, pulsar J1819+1305 exhibits the peculiar property 
of ``periodic nulls'', and Figure~\ref{fig4} gives a plot of pulse intensity 
{\it vs.} number which dramatically illustrates this strange phenomenon.  
Closer inspection, however, shows irregularities in the ``null'' spacings, 
and at several points the emission persists for up to about 100 pulses.  
Each of our PSs exhibits very strong ``null" modulation with a period of 
about 50 $P_1$, but of all our observations this MJD 53378 PS is the 
most regular.  Thus, it would seem that NAF's 1000-pulse observation 
was unusually periodic.  

Even more striking, perhaps, are the fluctuation spectra computed from 
the pulsar's PSs.  The longitude-resolved fluctuation (hereafter LRF) 
spectra in Figure~\ref{fig5}, in this case corresponding to the long MJD 
53778 observation, are typical of those resulting from our several 
observations.  All exhibit a primary fluctuation feature at roughly 0.017 
cycles/$P_1$ (hereafter c/$P_1$), corresponding to a period between 
50 and 60 $P_1$, which is seen across the full width of the profile.    
All of the harmonic resolved fluctuation (hereafter HRF) spectra have a 
structure like that of Figure~\ref{fig6}---that is, a pair of features at positive 
and negative frequencies of which the former is dominant.  A number of 
higher-frequency features are also visible in Figs.~\ref{fig5} and \ref{fig6} 
that will be further discussed in the next section. 

Accurate measurements of the low frequency modulation features were 
attempted across all our observations, and the results are summarized 
in Table 2.  In each case we used an HRF of length 256, and then 
examined the structure of the stronger (positive) feature at about 0.017 
c/$P_1$.  Two of our observations were sufficiently corrupted by RFI 
that no reliable determination could be made.  In the others the primary 
features were comprised of significant power in two Fourier components 
with periods of 51.2 and 64.0 $P_1$, and the tabulated values represent 
an appropriate weighting of their amplitudes.  Remarkably, the three 
reliable determinations yield compatible values of about 57$\pm$1 $P_1$.  
The large error reflects the limited facility of fluctuation spectra to determine 
the period of such a lengthy modulation cycle.  We can also probably 
understand how it was that NAF understood the nulls to have a roughly 
50-$P_1$ period.

That the star's ``periodic nulls'' give rise to such a well-defined low 
frequency modulation feature is itself a perplexing property.  They 
can be regarded, however, in a Fourier sense, as negative power.  
Apparently, we must thus conclude that the star's emission is strongly 
modulated with a period of about 57-$P_1$---and most significantly that 
this is so {\it in spite of} the obvious irregularities in the observed PSs 
(there evidently can be uninterrupted emission of up to 130 pulses, as 
in Fig.~\ref{fig9}).

\section{The emission structure of the bursts}
\begin{figure}
\begin{center}
\includegraphics[width=73mm,angle=-90]{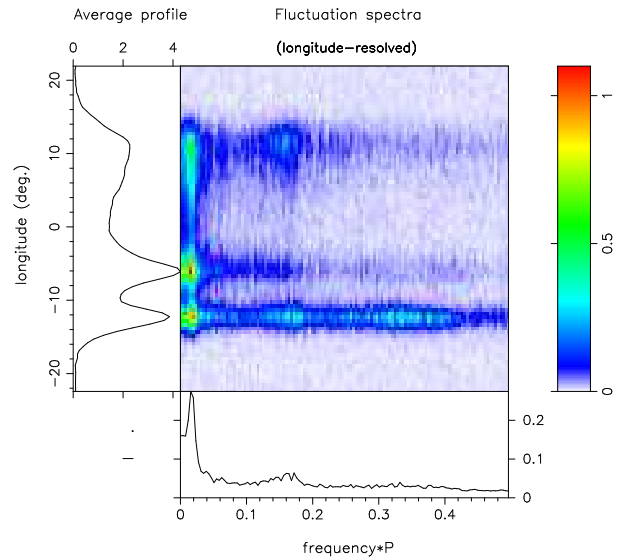}
\caption{Longitude-resolved fluctuation spectra of the 3394 pulses in 
the MJD 53778 observation. The prominent 50-$P_1$ feature is the 
dominant response in each of our PSs and is seen over the full width 
of the profile. However, features corresponding to $P_3$ values of 
$\approx$ 6.2 $P_1$ are prominent in the outer components, and 
evidence of weak $P_3 \approx 3$-$P_1$ modulation is exclusively 
present in the leading component.  A 256-length FFT was used to 
produce these spectra.}
\label{fig5}
\end{center}
\end{figure}

\begin{figure}
\begin{center}
\includegraphics[width=73mm,angle=-90]{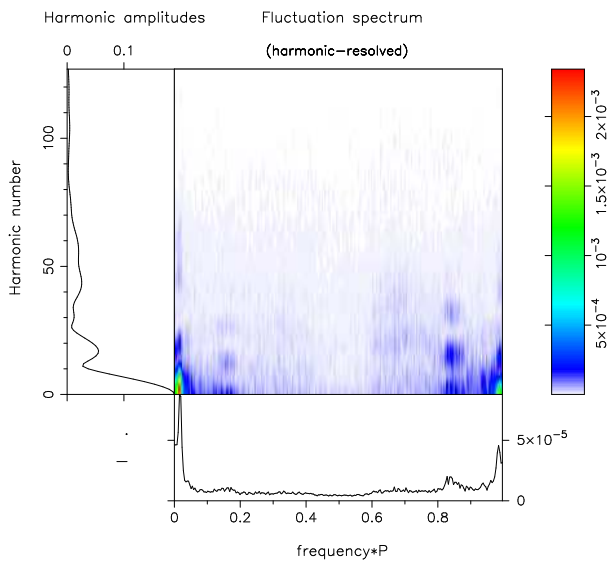}
\caption{HRF spectrum corresponding to the same PS as in Fig.~\ref{fig5}.  
Both the positive and negative frequency components corresponding to 
the 50-$P_1$ feature are evident, and the amplitude difference suggests 
that it represents a combination of amplitude and a phase modulation 
corresponding to  ``drift" from trailing-to-leading regions of the profile.  
By contrast, the asymmetric power at about 0.85 suggests leading-to-trailing 
``drift" of the 6.2-$P_1$ modulations seen in the spectrum of Fig.~\ref{fig5}}.
\label{fig6}
\end{center}
\end{figure}

We have established a picture of J1819+1305 as one of a series of 
emission bursts regularly punctuated by lengthy nulls. However, on 
inspection, the bursts themselves are found to exhibit characteristic 
emission patterns which are already hinted at by short-term features
in the spectra of Figs.~\ref{fig5} and ~\ref{fig6}. These features are 
found in all our observations, and are evidently not smeared out by 
long integrations. In Fig.~\ref{fig5} the two most obvious are the peaks 
corresponding to the periodicities of 6.2-$P_1$ and 3-$P_1$. The 
former is shared by the two outer components, suggesting a conal link, 
but the latter is confined to the first component. There is only weak evidence 
that the two inner components, presumably corresponding to a sightline 
intersection across an inner cone, show any regular short-term modulation. 
The HRF spectrum in Fig.~\ref{fig6}, having power corresponding to 
the two short-term features in its second half, indicates that these arise 
from positive drift (\ie, from leading to trailing longitudes), in contrast to 
the opposite weighting for the long-term modulation generated by the 
nulls. 

\begin{figure}
\begin{center}
\includegraphics[width=73mm,angle=-90]{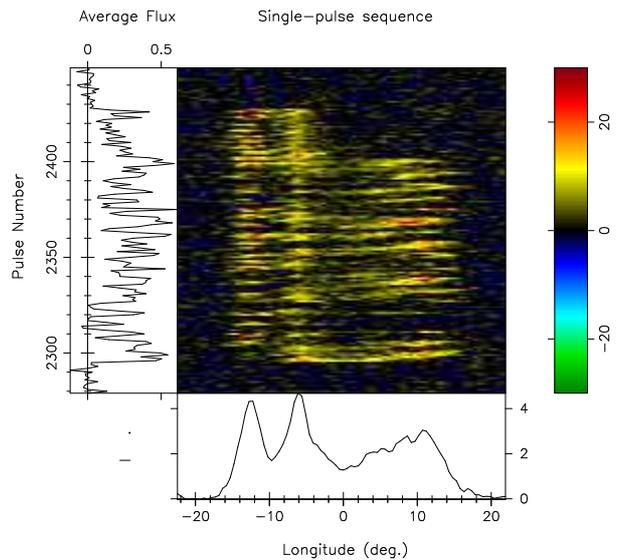}
\caption{Display of pulses 2281--2450 extracted from the MJD 53778 
observation. The PS is unusual in that it is not interrupted by nulls and 
also illustrates the character of both the 6.2- and 3-$P_1$ modulations.  
Note that the 6-$P_1$ pattern is retained in the trailing region after 
the 3-$P_1$ $P_3$ intermittently appears in the first component at 
about pulse 2340.  Note also the characteristic disappearance of the 
trailing components when the 3-$P_1$ pattern becomes dominant 
in the first component after about pulse 2405.}             
\label{fig9}
\end{center}
\end{figure}

Close inspection of the PSs reveals how these features interact in a 
two-pattern process.  Fig.~\ref{fig4} shows a series of typical bursts 
in the MJD 53378 observation, which can vary in length from 20 to 130 
pulses, and an exceptionally long burst from the MJD 53778 PS is 
plotted in detail in Figure~\ref{fig9}.  Most bursts, of whatever length, 
commence with emission in several components (always including
2), and the 6.2-$P_1$ modulation very evident as a clear positive 
drift in either or both of the outer components.   Then, after a number 
of 6.2-$P_1$ cycles, this modulation is rapidly transformed into a shorter 
sequence with the 3-$P_1$ modulation evident in the first component 
only.  In this second pattern, the third and fourth components are no 
longer detectable (\eg, after about pulse 2405 in Figure~\ref{fig9}).  The 
3-$P_1$ cycles are typically (but not invariably) repeated about the same 
number of times as the preceding longer cycles, so that this second 
behaviour lasts for less than half the duration of the first, before all 
emission ceases and the pulsar enters a null phase which may last 
anything between 10 and 40 rotation periods.  Short integrations over 
the two patterns of the bursts show that the initial phase has a broad 
4-peak intensity profile with no component dominating the others, 
whereas the second phase has just two sharp leading peaks. Clearly, 
the overall profile in Fig.~\ref{fig1} arises from the addition of these 
two modulation-pattern profiles. 

It can also be seen from Fig.~\ref{fig9} that the lack of emission in the 
latter part of the profile in the second burst pattern gives an overall 
intensity ``drift" behaviour from trailing-to-leading longitudes over 
the burst as a whole.  This is almost certainly the origin of the apparent 
asymmetry in the paired low-frequency features evident in the HRF of 
Fig.~\ref{fig6}. 

\begin{table}
\label{tab2}
\centering
 % \begin{minipage}{140mm}
 \caption{Pulse-Sequence Modulation Properties}
 \begin{tabular}{cccc}
 \hline
 Band  & MJD  &  Mod. Period  &Pulses  \\
%            &  Date &     $P_1/c$     &   &   \\
   \hline
   P      &       52707    &56.9.$\pm$1.0& 1697 \\
   P      &       52941    &           ---           &   955 \\
   P      &       53378    &57.3.$\pm$1.0&   848 \\
   P      &       53778    &57.0.$\pm$1.0&  3394 \\
   L      &       53859    &            ---          &   2075 \\
\hline
\end{tabular}
%\end{minipage}
\end{table}

The 6.2-$P_1$ periodicity may occur in either or both of the outer components. 
When the modulation is evident in both components there appears (by eye) 
to be a consistent delay of approximately 2 $P_1$ by which the trailing 
component's intensity follows that of the leading, but there is no evidence 
of drift bands linking the components right across the profile. However, in 
each of the outer components short positively-drifting bands can be seen. 
In the leading component the drift slows as the longitude of the second 
component is reached, a feature which is even more marked during the 
3-$P_1$ modulation: often the drift clearly becomes stationary at the 
second component, which therefore shows no significant modulation 
periodicity (Fig.~\ref{fig5}).  On those occasions when a driftband is clearly 
evident, especially in the trailing components, the drift-rate is slow.  If the 
drift were projected to match the modulated intensity peaks on the opposing outer 
components, drift bands would take about 8 $P_1$ to move across the full 
profile.
%Geoff, would you buy 11-P instead??

\begin{figure*}
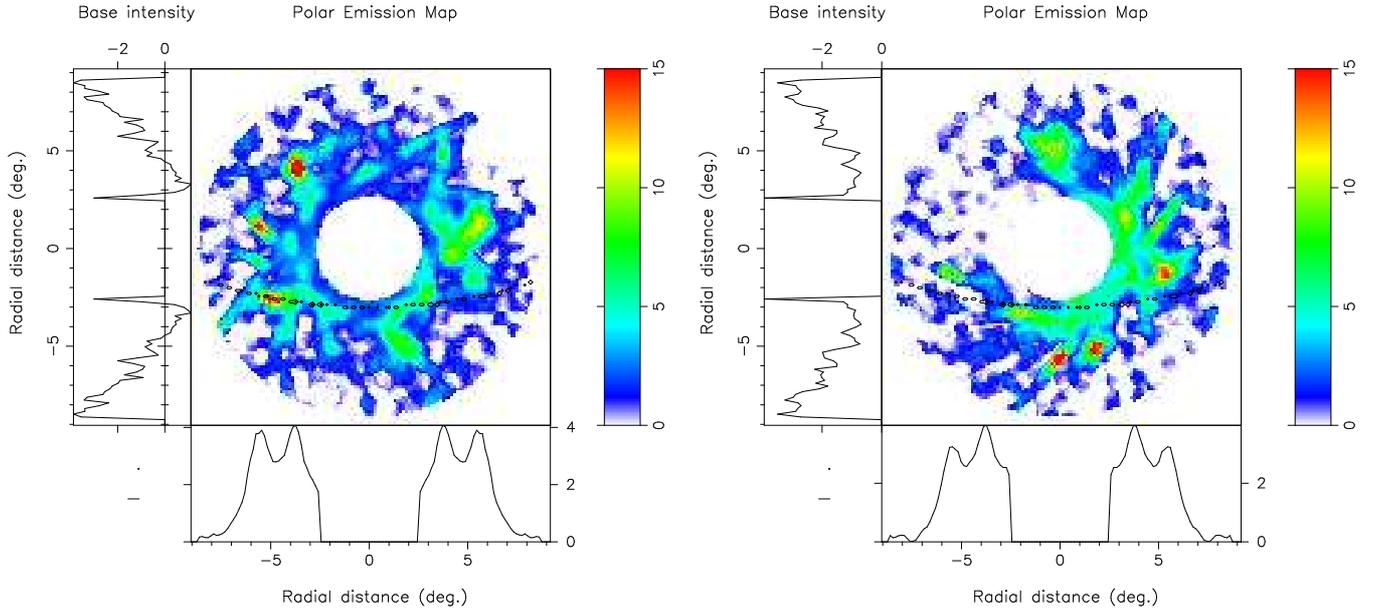

\begin{center}
\begin{tabular}{@{}lr@{}lr@{}}
{\mbox{\includegraphics[width=8cm,angle=-90]{MN-07-1367-MJr1_fig11a.ps}}}&
{\mbox{\includegraphics[width=8cm,angle=-90]{MN-07-1367-MJr1_fig11b.ps}}}\\
\end{tabular}
\caption{Subbeam ``carousel'' maps (central panels) of the MJD 53778 observation 
produced using the cartographic mapping methods described in Deshpande \& 
Rankin (2001).  Each is comprised of 57 pulses, the sequence beginning at pulse 
271 (left) has few ``nulls'' whereas that starting at pulse 2053 (right) includes a long 
``null'' interval.  The bottom panels give the radial emission profile and the side 
panels the unmodulated ``base'' (which has not been subtracted from the central 
emission patterns).  The sightline paths are shown in contours for an (assumed) 
outer traverse.}
\label{fig11}
\end{center}
\end{figure*}

It should be noted that there are frequent variations on this basic 2-pattern 
picture of the emission bursts. Sometimes one or other of the patterns is 
unusually long compared to the other. Occasionally, the first pattern
restarts rapidly after the second without an intervening long null sequence, 
leading to an unusually long emission burst.  The 3-$P_1$ modulation can 
seem sometimes to commence while the 6-$P_1$ modulation persists in 
the trailing components (as in Fig.~\ref{fig9} from about pulse 2340).
Nevertheless, there appears to be a ``rule", which must have some physical 
significance, that the 6.2-$P_1$ modulation is generally followed by the faster 
modulation, and not the reverse.  In this respect, J1819+1305 has much in 
common with B0031--07, B2319+60 and B1918+19, which also have 
``quantised" subpulse drift-rates governed by succession rules (see Wright 
\& Fowler 1981; Wright \& Fowler 1982; and Hankins \& Wolszczan 1987, 
respectively)

\section{Modeling the null--burst cycle}
Two basic properties of J1819+1305's subpulse behaviour have been 
established. Firstly, and primarily, that the null pulses tend to bunch 
together and that these bunches occur on a quasi-periodic basis.
Secondly, that the intervening bursts of emission have two characteristic 
patterns such that a 6.2-$P_1$ modulation in the outer components is 
usually followed by a 3-$P_1$ modulation present in the first component 
only.

Both these features suggest the idea of an underlying cycle regularly 
repeating itself, with secular modulations superimposed on the cycle so 
that no two cycles are the same. A natural model which would support 
these properties is that of the ``carousel'', whereby a pattern of ``beamlets'' 
are carried around the pulsar's magnetic axis close to its surface and are 
randomly sampled by the observer's line of sight. Such a carousel with, 
say, 70\% of its beamlets persistent and active would indeed generate 
a modulation feature reflecting its circulation time, as observed.  Whether 
or not one accepts the physical basis of this model (and it has been modified 
considerably over the years---\eg, Ruderman \& Sutherland 1975; Wright 
2003; Gil \etal 2006), in the present context its geometry is a convenient 
and dynamic way of modeling the observed cyclic patterns. 

In the carousel picture one naturally interprets the repetiton time of the cycle 
as a circulation time. This implies that the recurrent bunches of nulls are 
seen as regions of ``empty'' emission convected around the pole. It is then 
of interest to see how the observed emission bursts, with their dual modulation 
patterns and double cone structure, would appear in such a geometry.  We 
have therefore constructed visual images of the conjectured carousel---``polar 
maps'' using the cartographic transform of Deshpande \& Rankin (2001)---using 
the underlying geometry of the star, derived in \S 3, and assuming the mean 
circulation time of 57-$P_1$ derived in \S 4. The sense of the drift is taken as 
positive since this is the sense found in both the first and fourth components 
during the emission bursts.

\begin{figure}
\begin{center}
\includegraphics[width=73mm,angle=-90]{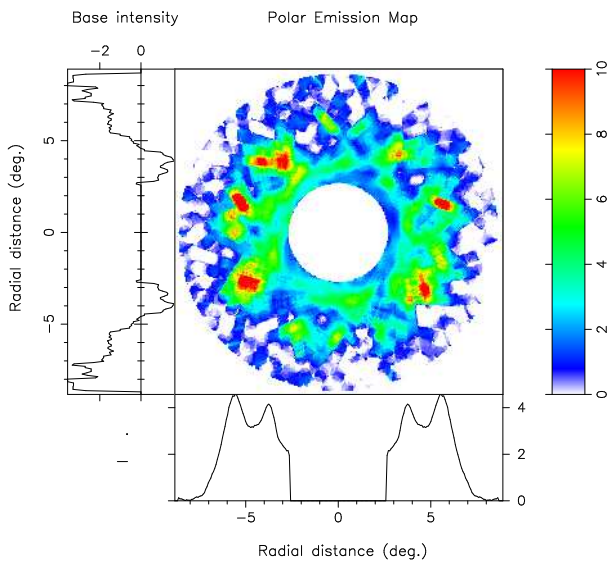}
\caption{Polar map as in Fig.~\ref{fig11} for the pulse 2296--2402 interval taken from
the MJD 53778 observation displayed in Fig.~\ref{fig9}.  This ``average'' map shows the 
subbeam-carousel structure responsible for the apparent subpulse drift 
with a $P_3$ of about 6.2 $P_1$.  }
\label{fig12}
\end{center}
\end{figure}

\subsection{Polar emission patterns}
We have computed polar emission maps and movies corresponding to the 
several available PSs, and two are shown in Figure~\ref{fig11}.  The left-hand 
display gives a polar map corresponding to an interval with relatively few 
``nulls'', whereas about half the pulses in the right-hand PS are ``nulls''.  We 
note that these maps correspond to only a single 57-$P_1$ circulation time, 
so they are necessarily more poorly sampled that those computed for stars 
with stable carousel configurations.  There is no evidence that most of the 
``beamlets'' encountered are active even over this short interval.  Indeed, 
when ``movies'' consisting of multiple successive frames like these are 
viewed together, we see little evidence that particular ``beamlets'' persist 
for a full circulation.  One can see instances in each map where the sightline 
traverses a given ``beamlet'' at several different angles, indicating that it 
persisted long enough to be tracked through both of the two outer cone 
components.  And other passes can be seen where only weak noise-like 
emission was encountered.  Also, we can see again here the double-cone 
structure of the map (and radial beamform in the lower panels) compatible 
with the emission geometry worked out in \S 3.  

Figure~\ref{fig12} gives a map corresponding to that part of the long burst 
of Fig.~\ref{fig9} where emission is seen across the entire profile (pulses 
2296-2402).  Here we see a few fairly regularly spaced ``beamlets'' which 
seem to produce the bright 6.2-$P_1$ LRF feature.  Note that these 
``beamlets'' are primarily seen in the outer conal region---and the feature 
is most prominent at these longitudes in the LRF---though there appears 
to be a weaker system in the inner cone as well.   Note also that a similar 
``beamlet'' pattern is seen in the sub-PS of Fig.~\ref{fig11} (left), and a closer 
inspection indicates that some of these ``beamlets'' do persist for a couple 
of circulation times at this point in the overall PS, perhaps explaining why 
the drift feature is more prominent here.  Finally, we have made ``movies'' 
consisting of multiple frames like those seen above.\footnote{These can be 
downloaded and/or viewed by accessing www.uvm.edu/~jmrankin under 
the ``Research/Images \& Movies'' menus.} 

We do not understand why it is that some pulsars exhibit a double conal 
emission configuration, nor indeed why most all conal emission has an 
either outer or inner conal geometry.  J1819+1305 and other pulsars with 
a similar configuration provide opportunity to examine the dynamics of the 
emission in the two cones.  That every measurable aspect of the ``beamlet''
structure in the two cones exhibits the same circulation time, drift direction, 
and $P_3$ strongly suggests that the emission within the respective cones 
is produced by the same set of emitting particles.

\section{Summary of results and discussion}
J1819+1305 exhibits an asymmetrical profile (Fig.~\ref{fig1}) over a broad 
band which appears to reflect emission from both the outer and inner emission 
cones.  Thus we have classified it as having a conal quadruple (cQ) profile 
despite the difficulty of distinguishing the trailing components in its profiles.  
Profile polarimetry permits us to estimate the star's magnetic latitude $\alpha$ 
and sightline impact angle $\beta$ as about 32\degr\ and 3.0\degr, respectively.  
$\beta/\rho$ should then be some 0.53. Though we have assumed an equatorward 
(positive) sightline traverse, no strong evidence exists to determine the sign of 
$\beta$.  

The pulse sequences (PS) of pulsar J1819+1305 were already known (NAF) 
to exhibit ``periodic nulls'', in which the pulsar's emission appears to switch on 
and off for several rotation periods with a fairly regular 50-rotation-period cycle. 
Our fluctuation spectra over multiple observations reveal that the star's PSs 
indeed regularly exhibit a bright and narrow low frequency feature.  This striking 
feature permits relatively accurate measurement and is found to represent a 
57$\pm$1-$P_1$ period in each observation---very nearly the same interval 
as seen qualitatively in the ``periodic nulls''.  Its positive and negative components 
show that it has both an amplitude and a phase-modulation character, reflecting 
the charactersitic patterns formed by the emission bursts which occur between 
the null bunches.    

Other than their periodicity, nothing about the ``nulls'' themselves appears 
extraordinary.  An analysis shows immediately that the populations of weak 
pulses and ``true'' nulls cannot be fully distinguished (Fig.~\ref{fig2}). However, 
a careful analysis of null and burst-lengths clarified the situation: while short nulls 
occurred randomly (as would typical weak pulses), longer nulls ($> 10$ periods)---
corresponding to the periodic bunches of ``true'' nulls---were over-abundant, 
suggesting some underlying structural feature. 

%reference Herfindal & Rankin

On examining the emission bursts between the null bunches we noted clear 
examples of subpulse drift toward increasing longitude (positive) at both the 
6.2- and 3-$P_1$ periodicities. The interaction of these two drift patterns 
is subtle.  The drift bands at the repetition rate of 6.2 $P_1$ traverse both the 
leading and trailing components.  By contrast, those of 3 $P_1$ are confined 
to the first component and are always accompanied by strong emission in the 
second component together with a striking lack of emission in the third and 
fourth components.  Furthermore, the 6-$P_1$ pattern always precedes 
3-$P_1$ intervals, although the 3-$P_1$ modulation sometimes begins 
before the 6-$P_1$ pattern expires.  Thus the pulsar's irregular aggregate 
profile is a combination of a regular four-peaked profile arising from the 6-$P_1$ 
modulation and the sharp two-component profile of the 3-$P_1$ modulation. 
The fact that there is a sequence rule between the two patterns in each 
burst, and therefore that emission in the trailing half of the profile ceases 
before that in the first naturally explains why the low frequency feature in the 
spectra suggests a long-term phase modulation towards the leading component. 
It also accounts for an observed strong anti-correlation between the leading 
and trailing components at a delay of around 22 $P_1$---approximately half 
the burst length.

%This motion appears to be slow, further suggesting 
%that it is unaliased---correlations, for instance, between the outer profile peaks are 
%greatest for a delay of 22 $P_1$.  Correlations also corroborate the star's double-cone 
%geometry; though its four components are not seen clearly in its profile, power in its 
%second component is correlated with the region around +6\degr\ longitude at a delay 
%of some 11 $P_1$.  Nonetheless, it remains unclear why the bright second component 
%is so well defined while its trailing counterpart is not so.  

J1819+1305's regular cycle of alternating long nulls and emission bursts can 
be modeled geometrically as a rotating carousel with the 57-$P_1$ modulation 
reflecting the carousel circulation time $\P3hat$, and the ``nulls'' representing 
sightline passes through the carousel which fail to encounter significant emission. 
This model therefore does not require nulls to correspond to a complete cessation 
of emission thoughtout the magnetosphere, merely that ``empty'' regions of the 
carousel circulate and maintain their identity for significantly longer than $\P3hat$, 
giving rise to a large and obvious population of pseudo-nulls.  We have argued 
earlier (Rankin \& Wright 2007) that the much smaller proportion of ``nulls'' in the 
Cambridge pulsar B0834+06 are also pseudo-nulls, and this second type of 
ostensible null may be more common that heretofore realized.  Herfindal \& Rankin 
(2007a,b), moreover, have identified null periodicities in Cambridge pulsar B1133+16 
and several other stars.  Can it be, for instance, that the nulls in pulsars B1918+19 
(Hankins \& Wolczszan 1987) and B1944+17 (Deich \etal\ 1986) are also of this 
character?

%The subpulse emission from both the star's cones appears to reflect precisely the 
%same circulation time, $P_3$ and drift direction, providing strong evidence that its 
%two cones are somehow produced by the same emission-carousel system.  This 
%conclusion pertains to the system as in the c{\bf Q} geometry different subbeams 
%in general produce the subpulses within the inner and outer cones which pass 
%our sightline.  

The carousel model, though here in essence no more than a geometric representation 
of the observed emission cycle, had its physical genesis in the work of Ruderman \& 
Sutherland's polar gap theory (1975).  In this respect it encounters difficulty.  Firstly, 
B1819+1305's $\P3hat$ value of 57 $P_1$ is very long compared to that expected 
in polar gap theory.  For a very usual pulsar with a 1.06-s $P_1$ and 6.3x$10^{11}$ 
G field, it would predict a value of barely 4 $P_1$.  Even if this star behaved as does 
B0943+10 (Deshpande \& Rankin 2001)---also with a longer value than predicted by 
the above theory---its $\P3hat$ would be some 12.6 $P_1$.  Thus to account for the 
much slower observed driftrate a reduced electric potential across the gap [as in the 
theoretical modifications suggested by Harding \etal\ (2002) or Gil \etal\ (2006)] may 
be needed.

Another perplexing difficulty results from the subtle behaviour of the inter-null 
emission within the cycle.  This shows two distinct drift modes, one about twice 
the repetition rate of the other.  Where this occurs in other pulsars (\eg, B0031--07, 
B2303+30, etc), it is easy to postulate that the circulation has, for some unknown 
reason, speeded up the very same ``beamlets'' (\eg Redman \etal 2005).  This is 
harder to allege for J1819+1305 because it would be in conflict with the 
ostensible regular carousel circulation rate.  We would have to reason that 
certain sectors of the carousel were twice as densely populated by ``beamlets'' 
than others, although this would not obviously explain why we tend to see 
the former immediately after the other.  Furthermore, the fact that the more rapid 
modulation at 3 $P_1$ is confined to the first component of the profile and 
generally accompanied by the ``nulling'' of the third and fourth components 
appears to coordinate two different sections of the underlying double cone, 
related only by the chance intersection of our line of sight.  One possibility is 
that the 3-$P_1$ modulation reflects a pattern of orthogonally polarised modal 
beamlets (as per Rankin \& Ramachandran 2003), but we have not been able 
to confirm this circumstance owing to the weakness of the pulsar's PS emission. 
However, this cannot easily explain why the 3 P1 modulation is {\em never}
found in the trailing components.

\section*{Acknowledgments}We thank Paulo Freire and Avinash Deshpande for 
bringing the peculiar properties of this pulsar to our attention. GW thanks the 
University of Sussex for a Visiting Research Fellowship.  Some of the work 
was made possible by US National Science Foundation Grants AST 99-87654 
and 00-98685.  Arecibo Observatory is operated by Cornell University under 
contract to the NSF.  This work made use of the NASA ADS astronomical data 
system.

{}

\bsp

\label{lastpage}

\end{document}